\documentclass[aps,prd]{revtex4}

\usepackage{latexsym}
\usepackage{amsfonts,amssymb}
\usepackage{graphicx,epsfig}
\usepackage{psfrag}

\newcommand{\be}{\begin{equation}}
\newcommand{\ee}{\end{equation}}
\def\bq{\begin{eqnarray}}
\def\eq{\end{eqnarray}}
\def\n{\nonumber}
\def\th{\theta}

\def\ph{\phi}

\def\tr{\overline{r}}
\def\S{\Sigma}
\def\x{\xi_0}

\begin{document}

\title{Spherical collapse with heat flow and without horizon}

\author{A.Banerjee\footnote{Correspondence to : asitb@cal3.vsnl.net.in},  S. Chatterjee\footnote{Permanent address: New Alipore College, Kolkata-700 053,India}}
\address{Relativity and Cosmology Research Centre, Deptt. of Physics, \\
Jadavpur University, Kolkata 700 032, India.}

\author{N. Dadhich}
\address{IUCAA, Post Bag 4, Ganeshkhind, Pune 411 007,India.}

\begin{abstract}
We  present  a  class  of solutions for a heat  conducting  fluid
sphere,  which radiates energy during collapse without the appearance of 
horizon at the boundary at any stage of the collapse. A  simple model
shows that there is no accumulation of energy due to collapse
since it radiates out at the same rate as it is being generated.
\end{abstract}

\maketitle

One of the interesting problems related to gravitational collapse
of stars is to study the solutions of various spherically
symmetric fluid distributions with radial heat flux in the
interior satisfying all the energy conditions and having
reasonable physical behaviour. In this case the exterior spacetime
is described by the radiating Vaidya metric [1]. The junction
conditions at the boundary of the isotropic fluid sphere with
dissipation in the form of the radial heat flux matching with the
radiating Vaidya metric were previously studied by Santos [2]. In
particular special attention was given to the model which from an
initial static configuration starts collapsing when dissipation
takes place in the form of heat flux [3]. The initial static
configuration evolves until the horizon is formed. \\
   
In this letter we present an interesting solution of a collapsing fluid sphere
with radial heat flux. The interesting feature of it is that collapse never 
encounters horizon and it continues until a (naked) singularity is reached. 
The matter content of this model has reasonable properties and the non-occurrence of the horizon is due to the fact that the rate of mass loss is balanced by the fall of boundary radius. In fact, one gets a class of such solutions although the explicit ex

ample is given for a very simple heat conducting fluid sphere, which satisfies all the energy conditions and has very reasonable physical properties. \\

The metric in the interior of a shearfree spherically symmetric fluid distribution is given by
\be
ds^2 = -A^2(r,t) dt^2 + B^2(r,t)[dr^2 + r^2 d \th^2 + r^2 \sin^2 \th d \phi^2]
\ee

The stress energy tensor of a nonviscous heat conducting fluid
reads
\be
T^{\mu \nu}  = (\rho + p) \, v^\mu v^ \nu + p g^{\mu \nu} + q^\mu v^\nu + q^\nu v^\mu.            
\ee
The heat flow vector $q^\mu$ is orthogonal to the velocity vector so
that $q^\mu v_\mu = 0$. Assuming comoving coordinates $v^\mu = (-g_{00})^{1/2} $ and $q^\mu = (0, q^1, 0, 0)$ nontrivial Einstein equations are [4]
\bq
\rho &=& - \frac{4 B'}{r B^3} + \frac{3 {\dot B}^2}{A^2 B^2} + \frac{B'^2}{B^4} - \frac{2 B''}{B^3} \\
p &=& \frac{2 A'}{r A B^2} + \frac{2 B'}{r B^3} + \frac{2 {\dot A} {\dot B}}{A^3 B} - \frac{{\dot B}^2}{A^2 B^2} - \frac{2 {\ddot B}}{B A^2} + \frac{2 A' B'}{A B^3} + \frac{{B'}^2}{B^4} \\
p &=& \frac{ A'}{r A B^2} + \frac{B'}{r B^3} + \frac{2 {\dot A} {\dot B}}{A^3 B} - \frac{{\dot B}^2}{A^2 B^2} - \frac{2 {\ddot B}}{B A^2} + \frac{A''}{A B} - \frac{{B'}^2}{B^4} + \frac{B''}{B^3} \\
q^1 &=& - \frac{2 A' {\dot B}}{A^2 B^3} - \frac{2 B' {\dot B} }{A B^4} + \frac{2 {\dot B}'}{A B^3}
\eq
 
The exterior Vaidya metric already mentioned is given explicitly in the form
\be
d s^2 = - \left(1 - \frac{2 M(v)}{\tr} \right) d v^2 - 2 d \tr d v + \tr^2 (d \th^2 + \sin^2 \th d\ph^2)
\ee
where v is the retarded time and M(v) is the exterior Vaidya mass.
The junction conditions then yield the following system of
equations valid on the boundary $r = r_{\S}$ [2],
\bq
(r B)_{\S} &=& \tr_{\S} \\
p_{\S} &=& (q^1 B)_{\S} 
\eq
and                  
\be
m_{\S} = \Bigg[\frac{r^3 B {\dot B}^2}{2 A^2} - r^2 B' - \frac{r^3 B'^2}{2 B} \Bigg]_{\S}
\ee
where $m_{\S}$  is the mass function
calculated in the interior at $r = r_{\S}$ [5, 6]. The junction
conditions yield $m_{\S} = M(v)$. \\

At this stage we consider a particular form of the metric coefficients given 
in (1) and choose them separable in r and t co-ordinates as
\bq   
A &=& a(r) \\           
B &=& b(r)R(t)    
\eq              
so that $\rho$ , p and $q^1$  can now be expressed in the form

\bq
\rho &=& \frac{1}{R^2} \bigg[\frac{3}{a^2} {\dot R}^2 - \frac{1}{b^2} \left(\frac{2 b''}{b} - \frac{b'^{2}}{b^2} + \frac{4 b'}{r b} \right) \bigg] \\
p &=& \frac{1}{R^2} \bigg[-\frac{1}{a^2} (2 R {\ddot R} + {\dot R}^2 ) + \frac{1}{b^2} \left(\frac{b'^2}{b^2} + \frac{2 a' b'}{a b} + \frac{2}{r} \left(\frac{a'}{a} + \frac{b'}{b} \right) \right) \bigg] \\
q^1 &=& - \frac{2 a' {\dot R}}{R^3 a^2 b^2}
\eq    
where `dot' and `dash' indicate derivatives with respect to time
and radial coordinate. The isotropy of pressure would give the equation,
\be
\frac{a''}{a} + \frac{b''}{b} - 2\frac{{b'}^2}{b^2} - 2\frac{a'b'}{ab} - \frac{a'}{ra} - \frac{b'}{rb} = 0.
\ee

The boundary condition (9) now yields at $r = r_{\S} = r_o$ in view of (14) 
and (15)
\be
2 R {\ddot R} + {\dot R}^2 + m {\dot R} = n
\ee
where m and n are constants. Such an equation was first given in [7] in the 
context of a very particular case of Maiti's solution [8]. The general solution of (16) in closed form is not available and a very simple solution is
\be
R(t) = - C t.      
\ee

An interesting feature of the above solution is that at the
boundary, $m_{\S}/ \tr{\S}$ turns out to be independent of time. It can be 
demonstrated
very easily from the expression (10). We then get using the
solution (18)
\be
\frac{2 m_{\S}}{\tr_{\S}} = \frac{2 m_{\S}}{(r B)_{\S}} = 2 \bigg[\frac{C^2 r_0^2 b_0^2}{2 a_0^2} - \frac{r_0 b_0'}{b_0} - \frac{r_0^2 b_0'^2}{2b_0^2} \bigg]
\ee
where $r_0$ is written for the comoving radial coordinate at the
boundary and $ b(r_0) = b_0, b'(r = r_0) = b_0'$. It is quite possible to 
adjust the aparameters in the above equation so as to keep $2 m_{\S} / \tr_{\S} < 1$  in order to avoid the appearance of horizon at the boundary. \\

{\bf A simple collapsing model with heat flow but without horizon} \\

We set $b(r) = 1$ in the pressure isotropy equation (16) and consider the special solution [9], 
\be     
A = a(r) = (1 + \xi_0 r^2).     
\ee
and then in view of equations (17-20) we get
\be   
m = - 4 \x r_0, \, \,   n = 4 \x  (1 + \x r_0^2),  \,\, 
C = \n \frac{1}{2} \bigg(-|m| + (m^2 + 4 n)^{1/2} \bigg).
\ee
In the solution $ R(t) = -C t$, the constant C is chosen to be
positive so that collapsing phase corresponds to $-\infty < t < 0$. We further have
\be
C^2 < 4 \x (1 + \x r_0^2) .
\ee
The explicit expressions for the density, pressure and heat flow
vector are now
\bq
\rho &=& \frac{3}{t^2 (1 + \x r^2)^2} \\
p &=& \frac{1}{t^2 (1 + \x r^2)^2} \bigg[\frac{4 \x}{C^2} (1 + \x r^2) - 1 \bigg] \\
q^1 &=& - \frac{4 \x r}{(1 + \x r^2)^2} \frac{1}{C^2 t^3}
\eq
All the above quantities diverge at $t \rightarrow 0$. The above expressions
show that $\rho > 0, p > 0$ and $\rho' < 0$, and $p' < 0$ would require
$C^2 < 2 \x (1 + \x r^2)$. It would be satisfied throughout the interior if 
$C^2 < 2 \x$. Further
$(\rho - p) > 0$ everywhere within the interior implies $C^2 >
\x(1 + \x r_0^2)$. All the above physically reasonable conditions will be
satisfied provided the following restrictions are imposed on the
magnitude of $C^2$
\be
2 \x > C^2 > \x (1 + \x r_0^2) .   
\ee
Since there is heat flow in the radial direction the fluid must
satisfy another condition in order to be consistent with all the
energy conditions and it is $(\rho + p) > 2|q|$, where $|q| \equiv
(g_{\mu \nu} q^{\mu} q^{\nu})^{1/2}$. This would require,
\be
1 + \frac{2 \x}{C^2} (1 + \x r^2) > \n  4 \x r/C 
\ee
which would always be true as it could be written as
\be
\Bigg[1 - \frac{2 \x r}{C} \Bigg]^2 > - \frac{2 \x}{C^2} (1 - \x r^2).
\ee \\

Now in view of eq. (19), since b = 1 it follows immediately that
\be  
1 - \frac{2 m_{\S}}{\tr_{\S}} = \Bigg[1 - \frac{C^2 r_0^2}{(1 + \x r_0^2)^2} \Bigg]           
\ee

So once we have $C^2 < 1/r_0^2 + \x^2 r_0^2 + 2 \x$, it is clear that the 
boundary surface can never reach the horizon. Here
in fact both the mass function and the area radius of the
radiating sphere vary linearly with t and hence the ratio is
independent of time. It may be further noted that though the
density, pressure and curvatures would diverge as $1/t^2$  when 
$t \rightarrow 0$, yet the mass
function which would go as t , would remain finite and go to zero
in this limit. Thus naked singularity would be a weak curvature singalarity. 
For strong
curvature  singularity the curvature  should diverge at least  as
$t^{-3}$. This simple  example  shows  that there is no accumulation of
energy  due to  collapse as it is  being radiated out at the same
rate as it is being generated.

{\bf Acknowledgement:} AB would like to thank IUCAA for hospitality.

\end{document}